\documentclass[11pt,fleqn]{cmes}
\cmes{1}{1}{2009}{0}

\graphicspath{{./Figures/}}
\runningtitle{Manuscript Preparation for CMES} 

\title{Discrete modelling of capillary mechanisms in multi-phase granular media}
\author{
 L. Scholt\`es\thanks{CSIRO Earth Science \& Resource Engineering - QCAT, Brisbane, Australia.},\
 B. Chareyre\thanks{Laboratoire Sols, Solides, Structures - Risques, Grenoble, France.}\
 F. Nicot\thanks{Cemagref - Unit\'e de recherche Erosion Torrentielle Neige et Avalanches, Grenoble, France.}\
and \ F. Darve\thanks{Laboratoire Sols, Solides, Structures - Risques, Grenoble, France.}\
}

\begin{document}
\maketitle

\abstract{
A numerical study of multi-phase granular materials based upon micro-mechanical modelling is proposed. Discrete element simulations are used to investigate capillary induced effects on the friction properties of a granular assembly in the pendular regime. Capillary forces are described at the local scale through the Young-Laplace equation and are superimposed to the standard dry particle interaction usually well simulated through an elastic-plastic relationship. Both effects of the pressure difference between liquid and gas phases and of the surface tension at the interface are integrated into the interaction model. Hydraulic hysteresis is accounted for based on the possible mechanism of formation and breakage of capillary menisci at contacts. In order to upscale the interparticular model, triaxial loading paths are simulated on a granular assembly and the results interpreted through the Mohr-Coulomb criterion. The micro-mechanical approach is validated with a capillary cohesion induced at the macroscopic scale.
It is shown that interparticular menisci contribute to the soil resistance by increasing normal forces at contacts. In addition, more than the capillary pressure level or the degree of saturation, our findings highlight the importance of the density number of liquid bonds on the overall behaviour of the material.
}
\keyword{Discrete Element Method, micromechanics, capillarity, multi-phase materials.}

\section{Introduction}
An outstanding feature of granular materials lies in the existence of different scales of interest. First, the microscopic scale corresponding to the scale of individual particles can be considered. The microscopic scale is intimately related to the interaction processes that occur between adjoining particles. Second, a mesoscopic scale can be identified, associated to a set of a few particles. The mesoscopic scale is relevant to describe local kinematics (\cite{Cambou2000}), or to address the creation and deletion of force chains amongst several elements (\cite{Radjai1999}).
Third, the macroscopic scale which is the scale of the specimen, generally denoted as the material scale.

As a matter of fact, the overall mechanical behaviour of granular materials is intimately related to the local properties that take place at the microscopic scale. In addition to dry contact interactions, the presence of a surrounding liquid can modify local deformation, attrition, and sliding between particles. The material is thus a tri-phasic medium where the skeleton, liquid and air interact depending on the thermodynamic equilibrium. According to the liquid content and to the void ratio of the medium, different saturation regimes can be observed. For small liquid contents, as long as the liquid tends to be concentrated as independent menisci between adjoining particles, this is the pendular regime. For increasing liquid content, menisci begin to merge. From this point, all liquid-gas configurations until the gaseous phase consists in gas bubbles within the liquid are classified as funicular configurations. Beyond, the medium is in a capillary state until it tends toward a fully saturated regime. For each case, the presence of both liquid and gas induces attractive forces that confer specific constitutive properties to the material. However, even though interparticular menisci have been dealt with a great attention in the pendular regime (\cite{Hotta1974, Lian1993, Soulie2006}), their modelling for higher saturation regimes is still a great challenge to deal with due to complex thermodynamic arrangements between the three phases (\cite{UrsoLawrence1999, Murase2004}). The present paper focuses on liquid bridges as defined in the pendular state as an attempt toward the understanding of induced complex macroscopic behaviours. The analysis is therefore limited to low saturation degrees where the pendular assumption can be assumed.

Assessing the constitutive behaviour of such multi-phasic materials at the macroscopic scale can be addressed through sophisticated phenomenological constitutive models, requiring specific assumptions and mathematical refinements (see \cite{Cambou1998} for a general review). This line of thinking has prevailed during the past decades. A powerful alternative concerns the micromechanical approaches, in which the description of physical phenomena is dealt with at the microscopic scale. The overall behaviour is therefore derived thanks to homogeneous schemes, relating both microscopic and macroscopic scales as presented in \cite{Christoffersen1981, Cambou1998, Kruyt2000, Nicot2005} or \cite{Lu2006} for example. 
Such methods are appealing since the local physics can be described by simple equations, requiring a small number of parameters with generally a clear physical meaning.

Following a micromechanical approach, discrete element methods (DEM) constitute a particularly relevant computational tool to track the constitutive response of granular masses along a variety of loading paths. Indeed, the rheological behaviour is obtained without any global hypotheses. Since the pioneering work of \cite{Cundall1979}, DEM approach has been popularized in the field of granular materials, and extension to multi-physical coupling issues is an emergent line of research. For example, hydro-mechanical modelling supports now a great source of efforts, both for unsaturated (\cite{Richefeu2007, ElShamy2008, Scholtes2008}) and saturated flows (\cite{Han2007, ElShamy2008b}). This paper is situated in this context, by showing how the local parameters characterizing wetted contact interactions within a granular medium can influence the overall response of the material. In the following, the numerical specimen is assumed to be homogeneous and constituted by a sufficient number of grains so as a constitutive relation can be derived from both homogeneous strain and stress tensors: the specimen is considered as a Representative Volume Element (RVE) of a granular medium

\section{Micromechanical model}
DEM has been extensively used to study soil mechanics providing, for instance, some insights into shear strength and deformation properties of granular soils (see \cite{Iwashita1988} and \cite{Ting1989} for example). Recently, DEM has been enhanced to investigate unsaturated granulates features by considering the possible effects of capillary water between grains (\cite{GiliAlonso2002, Jiang2004, Richefeu2006, ElShamy2008, Scholtes2008}). The DEM is essentially a Lagrangian (mesh-free) technique where each particle of the material is a sphere identified by its own mass, radius and moment of inertia. For every time step of the computation, interaction forces between particles, and consequently resulting forces acting on each of them, are deduced from sphere positions through the interaction law. Newton's second law is then integrated through an explicit second-order finite difference scheme to compute new spheres' positions.

We present here a 3D discrete element model developed into the YADE-Open DEM platform (\cite{Kozicki2008}), where effects of capillarity have been implemented in both terms of force and water retention through the resolution of the Young-Laplace equation as presented hereafter. In the proposed model, each interaction can thus be described as the sum of a dry contact repulsive force and of an attractive capillary force resulting from the presence of liquid bridges between grains.

\subsection{Contact friction model}
\begin{figure}
\centering
\includegraphics[width=.6\linewidth]{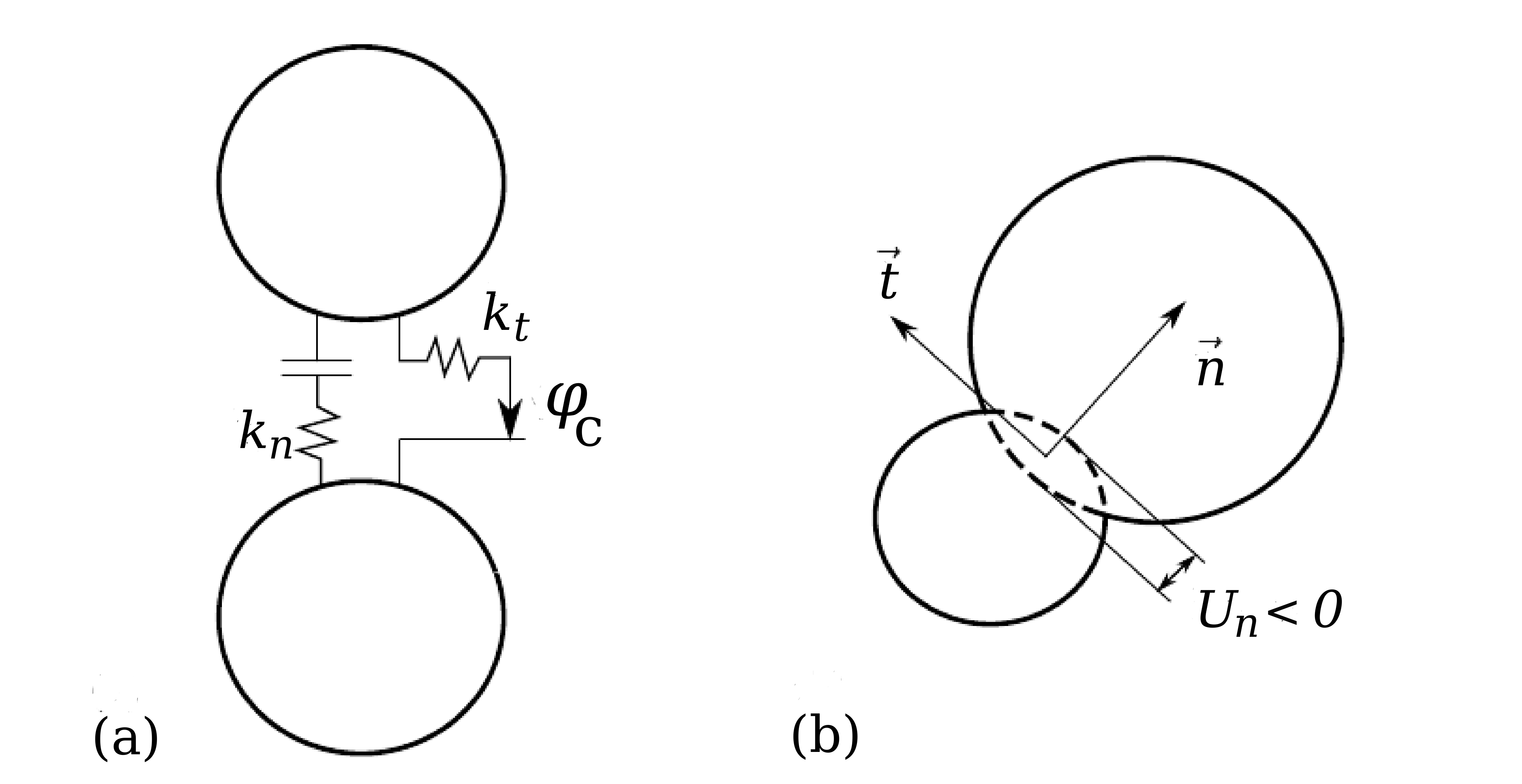}
\caption{Contact interaction: (a) model and (b) overlap.}
\label{ContactModel}
\end{figure}

The elastic-plastic model for contact friction is presented in \fig{ContactModel}(a). Although the formulation is quite simplistic compared to more realistic Hertz-Mindlin solutions, it has been proved to give accurate results (\cite{DiRenzo2003}), with a significant computational efficiency.

\begin{figure}
\centering
\includegraphics[width=.8\linewidth]{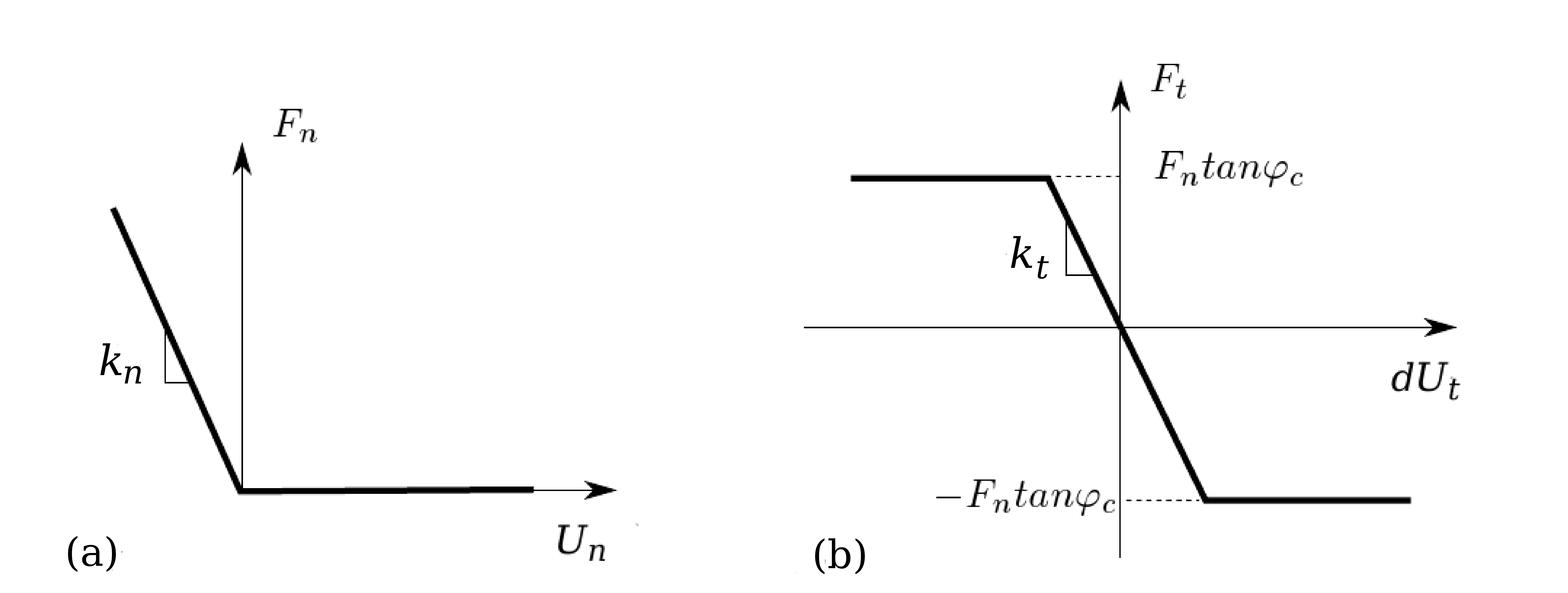}
\caption{Contact friction law: (a) normal force and (b) tangential force.}
\label{ContactLaw}
\end{figure}

Particles are considered to be rigid, but can overlap as shown in \fig{ContactModel}(b). This overlap accounts for the deformation induced by contacts forces. A linear elastic law provides the contact force as a function of the relative displacement between two interacting grains (see \fig{ContactLaw}). A normal stiffness $K_n$ is defined to relate the normal force $F_n$ to the intergranular normal distance $U_n$ such as:
\begin{equation}
F_n = \left\{ \begin{array}{ll} K_n U_n , \quad \textrm{if} \quad U_n \leqslant 0\\
				0 , \quad \textrm{if} \quad U_n > 0
	      \end{array} \right.
\label{Kn}
\end{equation}
and a tangential stiffness $K_t$ allows to deduce the shear force $F_t$ induced by the incremental tangential relative displacement $dU_t$:
\begin{equation}
dF_t = -K_t dU_t
\label{Kt}
\end{equation}
$K_n$ and $K_t$ are dependent functions of the interacting particle radii $R1$ and $R2$ and of a characteristic modulus $E$ of the material such as:
\begin{equation}
\left\{ \begin{array}{ll} K_n = 2.\frac{E.R1.R2}{(R1+R2)}\\ 
			  K_t = \alpha.K_n	\end{array} \right.
\label{E}
\end{equation}
with $\alpha$ a fixed parameter. This definition results in a constant ratio between $E$ and the effective bulk modulus of the packing, whatever the size of the particles.\\

Shear and normal forces are finally related by a slip Coulomb model such that $F_t^{max} = -\mu F_n$, where $\mu$ is the contact friction coefficient defined as $\mu = tan(\varphi_c)$, with $\varphi_c$ the intergranular friction angle.

\subsection{Capillary model}
\begin{figure}
\centering
\includegraphics[width=\linewidth]{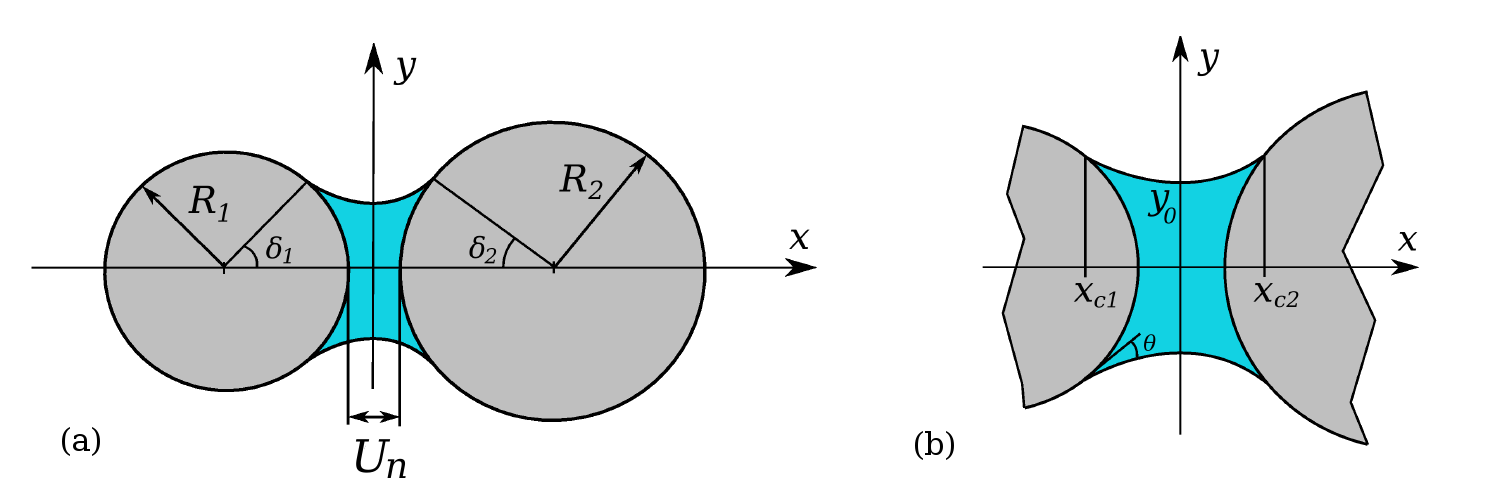}
\caption{Illustration of a liquid bridge between two particles of unequal sizes: (a) global geometry, (b) details of the bridge.}
\label{Doublet}
\end{figure}

For simplicity, we assume that capillary water inside the sample is solely composed of interparticular independent menisci as defined in the pendular state (\fig{Doublet}).

Their exact shape between spherical bodies is defined by the Young-Laplace equation, which relates the pressure difference $\Delta u = u_g - u_l$  across the gas-liquid interface to the radius of curvature $C$ of the bridge surface and to the surface tension $\gamma$ of the liquid phase:
\begin{equation}
\Delta u = \frac{\gamma}{C}
\label{LaplaceBasic}
\end{equation}

In the Cartesian coordinates of \fig{Doublet}, $C$ can be formulated as a function of the profile $y(x)$ of the liquid-gas interface curve, the $x$ axis coinciding with the axis of symmetry of the bridge, passing through the centers of the bonded spheres. Following the Young-Laplace equation (\eq{LaplaceBasic}), the profile of the liquid bridge is thus related to the capillary pressure $\Delta u$ through the following non-linear differential equation:
\begin{equation}
\frac{\Delta u}{\gamma}(1+y'^2(x))^{3/2} + \frac{1+y'^2(x)}{y(x)} - y''(x) = 0
\label{Laplace}
\end{equation}
According to a recent study by \cite{Soulie2006}, the corresponding meniscus volume $V_m$ and intergranular distance $U_n$ can be obtained by considering the $x$-coordinates ($x_{c1}$ and $x_{c2}$) of the three-phases contact lines defining the solid-liquid-gas interface such as:
\begin{equation}
\begin{array}{cc}
V_m = \pi \int_{x_{c1}}^{x_{c2}} y^2(x)dx - \frac{1}{3} \pi R_1^3 (1-acos(x_{c1}))^2(2+acos(x_{c1}))\\
 - \frac{1}{3} \pi R_2^3 (1-acos(x_{c2}))^2(2+acos(x_{c2}))
\end{array}
\label{MeniscusVolume}
\end{equation}
and
\begin{equation}
U_n = R_2(1-acos(x_{c2}))+x_{c2}+R_1(1-acos(x_{c1}))-x_{c1}
\label{IntergranularDistance}
\end{equation}

The capillary force due to the liquid bridge can then be calculated at the profile apex $y_0$ according to the `gorge method' (\cite{Hotta1974}) and consists of a contribution of both $\Delta u$ and $\gamma$:
\begin{equation}
F_{cap} = 2\pi y_0\gamma + \pi y_0^2\Delta u
\label{CapillaryForce}
\end{equation}

The relation between the capillary pressure and the configuration of the capillary doublet is thus described by a system of non-linear coupled equations (\eq{Laplace}, \eq{MeniscusVolume}, \eq{IntergranularDistance}, \eq{CapillaryForce}) where local geometry ($U_n$) and meniscus volume ($V_m$) arise as a result of the solved system. 

\begin{figure}
\centering
\includegraphics[width=.6\linewidth]{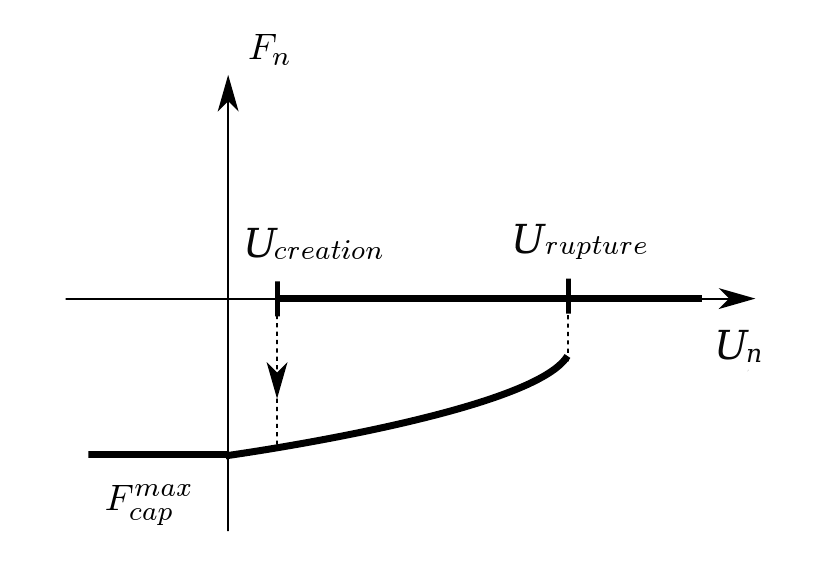}
\caption{Evolution of the capillary force with the relative displacement between interacting grains for a given capilary pressure $\Delta u$: a meniscus can form for $U_n \leq U_{creation}$ and breaks for $U_n > U_{rupture}$.}
\label{CapillaryLaw}
\end{figure}

In order to account for capillarity in the YADE-Open DEM code, an interpolation scheme on a set of discrete solutions of the Young-Laplace equation has been developed so as to link directly $\Delta u$ to $F_{cap}$ and $V_m$ for a given grain-pair configuration ($R_1$, $R_2$, $U_n$). This numerical procedure results in a suction-controlled model where, at every time-step during the simulation, capillary forces and water volumes are computed based upon the local geometry and the imposed capillary pressure. \fig{CapillaryLaw} presents the evolution of the capillary force with the relative displacement between two interacting grains.

\begin{figure}
\centering
\includegraphics[width=0.8\linewidth]{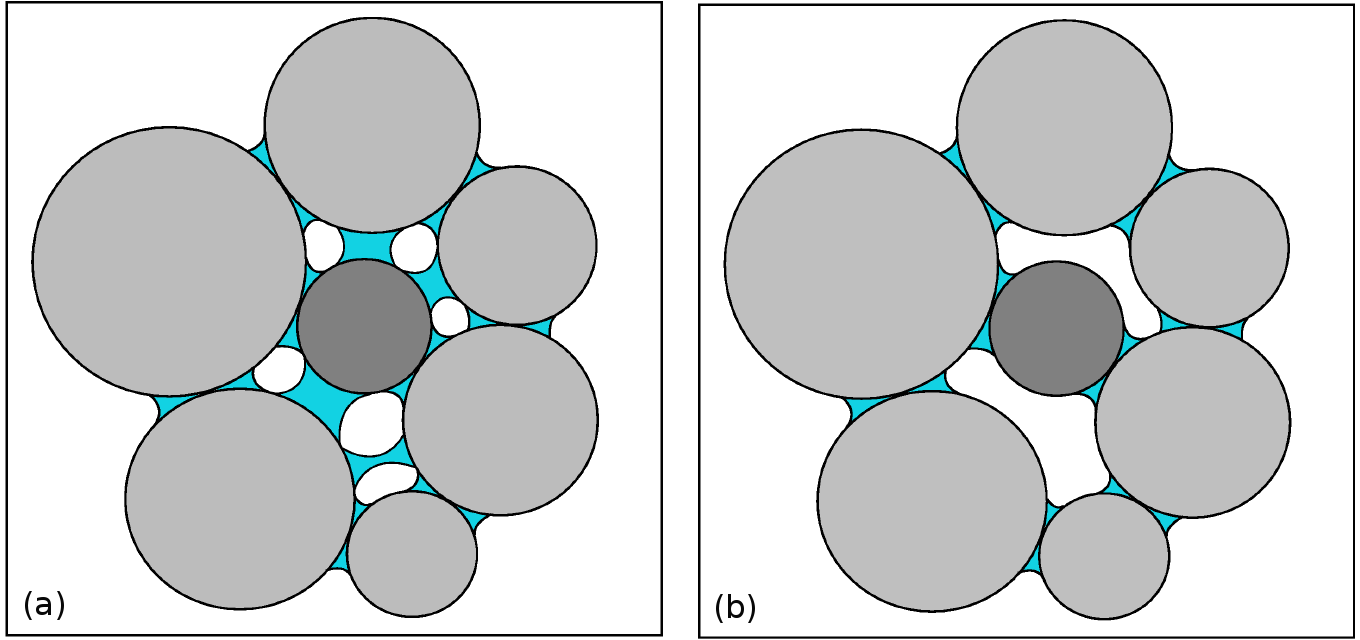}
\caption{Hydraulic hysteresis history accounted for in the code: assembly of particles resulting from (a) a drying scenario, (b) a wetting scenario.}
\label{HysteresisScheme}
\end{figure}

It is to be noted that capillary forces act exclusively in the axial direction of the interaction (the normal to the tangential contact plane), and that $F_{cap}$ is maximum for grains in contact ($U_n \leqslant 0$). $U_{rupture}$ is the debonding distance corresponding to the minimum $U_n$ value from which the Young-Laplace equation has no solution. An hydraulic hysteresis can be accounted for by defining $U_{creation}$ as the distance from which a liquid bridge can form between two interacting grains. For instance, as a drying scenario results in a desaturation of the material, liquid bridges can exist at stretched contacts as long as the debonding distance $U_n$ is less than the critical separation distance $U_{rupture}$ (\fig{HysteresisScheme}(a)). On the other hand, a wetting scenario only allows appearance of menisci between contacting or close grains (\fig{HysteresisScheme}(b)) such as it would occur during a capillary condensation when the relative humidity of the surrounding air is increased. The choice was made here to allow bridge appearance for strict contact ($U_{creation} = 0$), thus neglecting the possible effect of adsorbed water. In the same way, once they break, menisci can reform when particles come strictly in contact.

Following these assumptions, wetting and drying scenarios can be taken into
consideration for the initial state of the material, keeping in mind that, in
our simulations, the contact angle $\theta$ (\fig{Doublet}(b)) which defines the
wetting of the material is set to 0$^{\circ}$ for simplicity, assuming a perfect
wetting of the material and occulting therefore its possible effect on local
hysteresis mechanisms. 

\subsection{Interparticle behaviour}
The  interaction laws involve normal repulsion and Coulomb friction at contact
as well as capillary adhesion as presented respectively in \fig{ContactLaw} and
\fig{CapillaryLaw}.

\begin{figure}
\centering
\includegraphics[width=\linewidth]{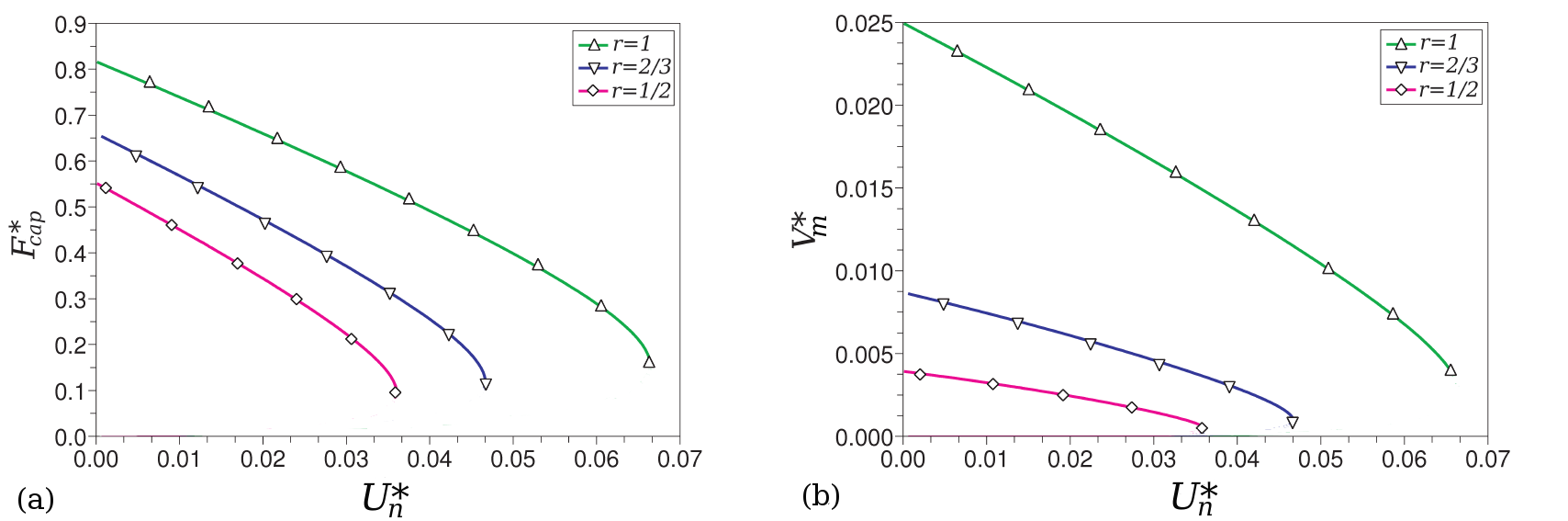}
\caption{Effect of the particle size ratio $r = \frac{R_1}{R_2}$ on (a) the normalized capillary force $F^*_{cap} = \frac{F_{cap}}{2 \pi R_2 \gamma}$ and (b) the normalized meniscus volume $V_m^* = \frac{V_m}{R_2^3}$ as a function of the separation distance $U_n^* = \frac{U_n}{R_2}$ for $\Delta u = 1000$~kPa.}
\label{ParticleSizeInfluence}
\end{figure}

Capillary forces are considered constant for the range of elastic contact deformation ($U_n \leqslant 0$), assuming a low deformability of particles during simulations. To pull two bonded spheres apart, an increasing external tensile force is needed. This force has to balanced the sum of the repulsive contact force $F_n$ and the adhesive capillary one $F_{cap}$ which tends to decrease with the interparticle distance $U_n$. The tensile process is stable as long as $U_n$ is less than the debonding distance $U_{rupture}$ which will depend on both the doublet configuration ($R_1$, $R_2$) and the capillary pressure $\Delta u$. As presented in \fig{ParticleSizeInfluence}, the particle size ratio $r = \frac{R_1}{R_2}$ has a great influence on both the adhesive force and the liquid bridge volume. For instance, for a given $\Delta u$ value, the greater the ratio $r$ is, the smaller the bridge volume $V_m$ is, resulting in less significant interacting distances $U_{rupture}$ before rupture as well as in diminishing force intensities.

\begin{figure}
\centering
\includegraphics[width=\linewidth]{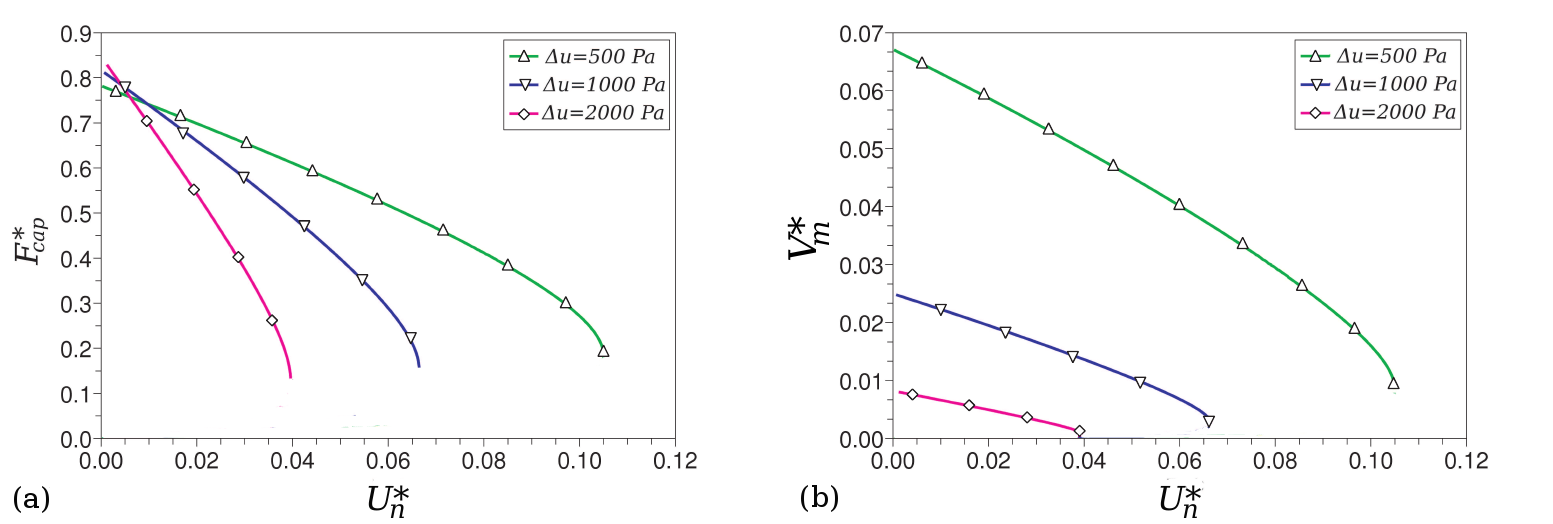}
\caption{Effect of the capillary pressure $\Delta u$ on (a) the normalized capillary force $F^*_{cap} = \frac{F_{cap}}{2 \pi R_2 \gamma}$ and (b) the normalized meniscus volume $V_m^* = \frac{V_m}{R_2^3}$ as a function of the separation distance $U_n^* = \frac{U_n}{R_2}$ for $r = \frac{R_1}{R_2} = 1$.}
\label{CapillaryPressureInfluence}
\end{figure}

Concerning now the influence of the capillary pressure (\fig{CapillaryPressureInfluence}), it has to be noted that, for a given doublet configuration ($r=1$), if higher values of $\Delta u$ result in smaller liquid bridge volumes, capillary force intensity at contact ($U_n = 0$) is quite constant whatever $\Delta u$. At the particle scale, capillary pressure, through its effect on menisci volumes, finally influences more the debonding distance than the resulting adhesive force.

To sum up, in addition to the great influence of particle size distribution on their intensity level, capillary effects between interacting grains are strongly driven by the amount of liquid involved in their behaviour. Such a complexity leads to numerous uncertainties when considering granular materials such as soils. The next section is therefore an attempt to provide some clarifications on capillarity consequences at the scale of a numerical granular assembly.

\section{Macroscopical behaviour}
\begin{figure}
\centering
\includegraphics[width=0.5\linewidth]{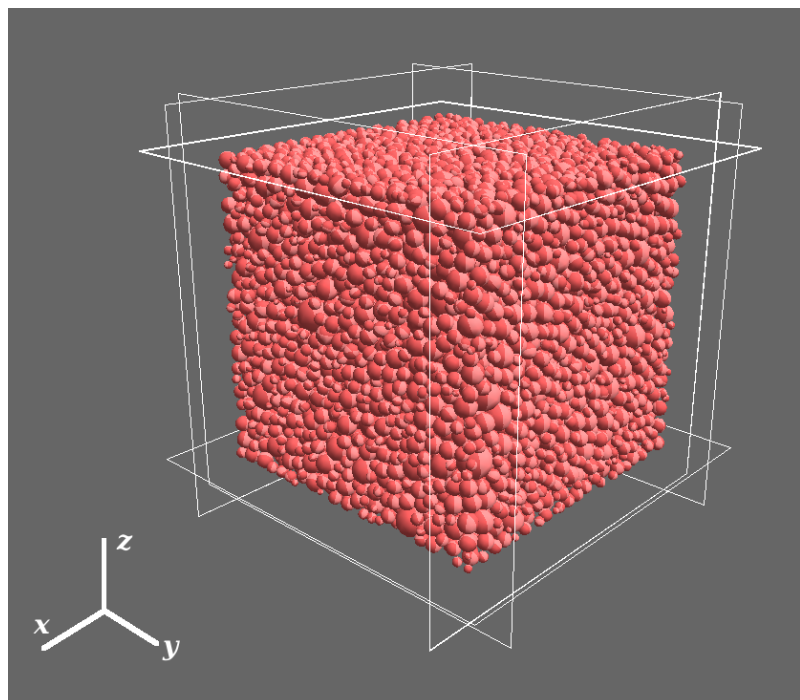}
\caption{The discrete element model.}
\label{Assembly}
\end{figure}

In order to investigate macroscopic consequences of interparticle capillary menisci, numerical simulations were conducted on a polydisperse assembly composed of 10,000 spheres (\fig{Assembly}) with a uniform grain size distribution ranging from 0.035~mm to 0.07~mm. The specimen is contained inside a box made of six rigid frictionless boundary walls for which positions at each time steps are defined from the prescribed loading program. \tab{Input} provides a summary of the input parameters used in the simulations, referring to \eq{Kn}, \eq{Kt} and \eq{E}.
\begin{table}
\centering
\caption{Input parameters}\label{Input}
\vskip1mm
\begin{tabular}{|c|c|c|}
\hline
Global Modulus	& $\frac{k_t}{k_n}$	& Friction angle\\
$E$ (MPa)	& $\alpha$		& $\varphi_c$ (deg.)  \\
\hline
50		& 0.5			& 30 \\
\hline
\end{tabular}
\end{table}

The sample is prepared in a dry configuration by an isotropic-compaction
technique which ensures the initial homogeneity of the packing
(\cite{Scholtes2008}). The final porosity of the sample depends on the
intergranular friction angle $\varphi_c$ defined during the compaction stage.
The smaller $\varphi_c$, the denser the specimen. Here, $\varphi_c$ was fixed to
1$^{\circ}$, leading to a dense specimen with a porosity of about 0.39.
$\varphi_c$ was then set to 30$^{\circ}$ for all the simulations. Starting from
the stabilized isotropic configuration, capillary pressure $\Delta u$ is then
switched on, letting menisci to be formed between all possible interacting
grains following the chosen drying or wetting scenario (stretched or contacting
menisci respectively). It is to be noted that the suction controlled scheme
ensures menisci to be homogeneously distributed inside the medium in accordance
the thermodynamic equilibrium between both liquid and gas phases.

\subsection{Numerical results}
Several loading paths were applied to the specimen in order to assess both its hydraulic and mechanical properties.

\subsubsection{Hydraulic properties}
Determining hydraulic properties of unsaturated soils is of great interest because of the strong hydro-mechanical coupling consequences on the overall behaviour. Herein, the proposed model allows to determine the water content of an assembly as a result of the summation from all menisci volumes contained inside the medium. Indeed, the degree of saturation $Sr$ of the numerical sample is obtained by:
\begin{equation}
Sr = 100*\frac{V_l}{V_v} = 100*\frac{\sum^{N_m}_{m=1} V_m}{V_{sample} - V_{grains}}
\label{SaturationDegree}
\end{equation}
with $V_l$ the total liquid volume and $V_v$ the void volume resulting from the grains arrangement inside the cubic box. $V_v$ is equal to the difference between the total volume of the sample $V_{sample}$ minus the volume occupied by all the grains $g$ forming the assembly ($V_{grains} = \sum^{N_g}_{g=1} \frac{4}{3} \pi R_g^3$). By varying the capillary pressure $\Delta u$ inside a given sample, one can therefore construct its soil water characteristic curve (SWCC).

However, due to the pendular assumption concerning liquid phase modelling (independent interparticular liquid bridges), the SWCC cannot be computed for high degrees of saturation $Sr$. Indeed, below a given capillary pressure, liquid bridges become big enough to allow their overlap with neighbours. As this overlap leads to menisci fusion associated with a complete fluid reorganisation which makes obsolete the pendular assumption, a numerical procedure has been developed in order to identify menisci superpositions through the definition of their geometrical configuration. As defined in \fig{Doublet}, the wetting of a liquid bridge over the grain surface is defined by its filling angles $\delta_1$ and $\delta_2$. It is therefore possible to determine if, for a given grain $g$ of an assembly, neighbouring menisci meet through their wetted surfaces.

\begin{figure}
\centering
\includegraphics[width=\linewidth]{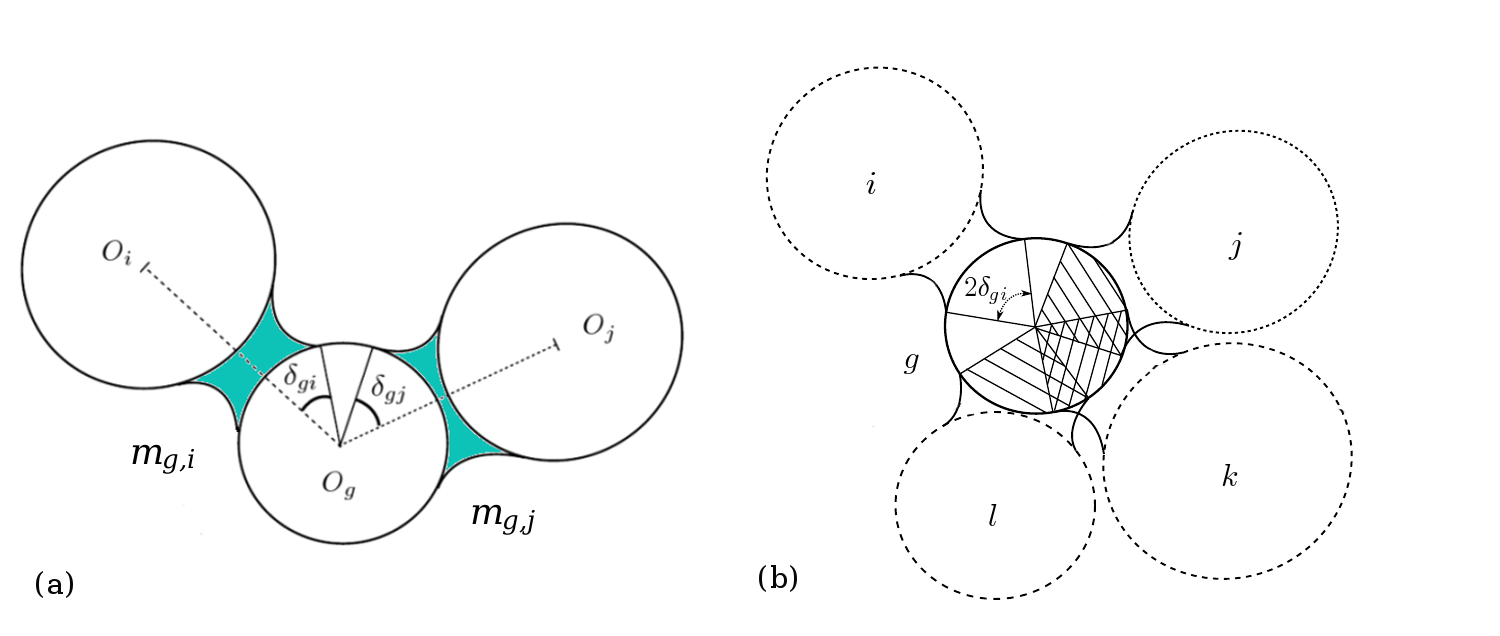}
\caption{illustration of the numerical procedure for menisci fusion: (a)
definition of the local geometry, (b) identification of merged menisci:
$m_{g,j}$, $m_{g,k}$ and $m_{g,l}$ overlap on grain $g$ surface.}
\label{FusionMenisci}
\end{figure}

For instance (see \fig{FusionMenisci}), if the angle $\widehat{O_i O_g O_j}$ formed between three interacting grains $g$, $i$ and $j$ is smaller than the sum $(\delta_{gi} + \delta_{gj})$ corresponding to the surface of the grain wetted by menisci $m_{g,i}$, and $m_{g,j}$ linking $g$ with $i$ and $g$ with $j$ respectively, an overlap occurs and fusion can therefore process. Obviously, the occurence of menisci fusion strongly depends on the microstructural arrangement and on the porosity of the medium. The range of saturation that corresponds to the pendular regime cannot therefore be infered for all materials and has to be checked for each particular case.

\begin{figure}
\centering
\includegraphics[width=0.6\linewidth]{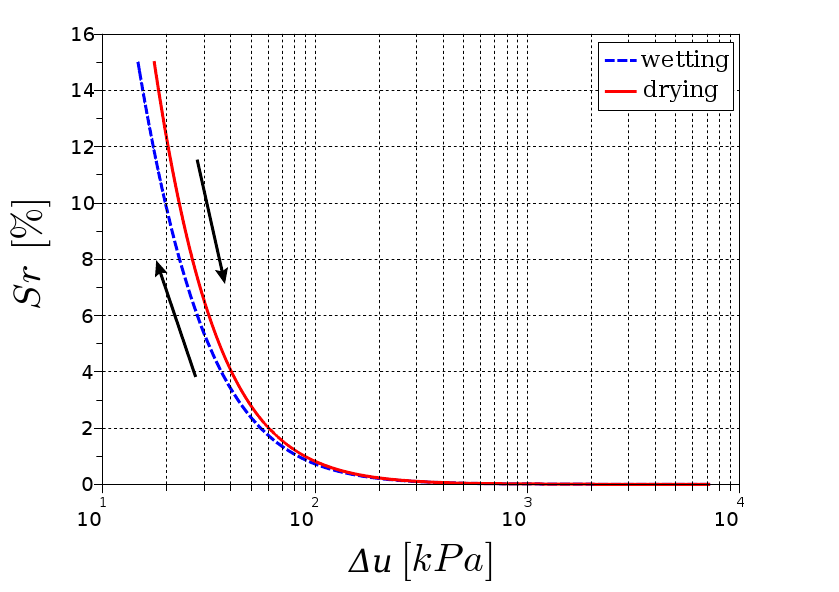}
\caption{Soil Water Characteristic Curve generated with the proposed discrete element model.}
\label{CharacteristicCurve}
\end{figure}

\fig{CharacteristicCurve} presents the SWCC obtained for the predefined numerical assembly in the range of its pendular regime. For this particular case, our finding is that the independence of menisci is strictly ensured for $Sr < 12\,\%$, whereas overlappings are negligible until $Sr = 15~\%$.

\begin{figure}
\centering
\includegraphics[width=0.6\linewidth]{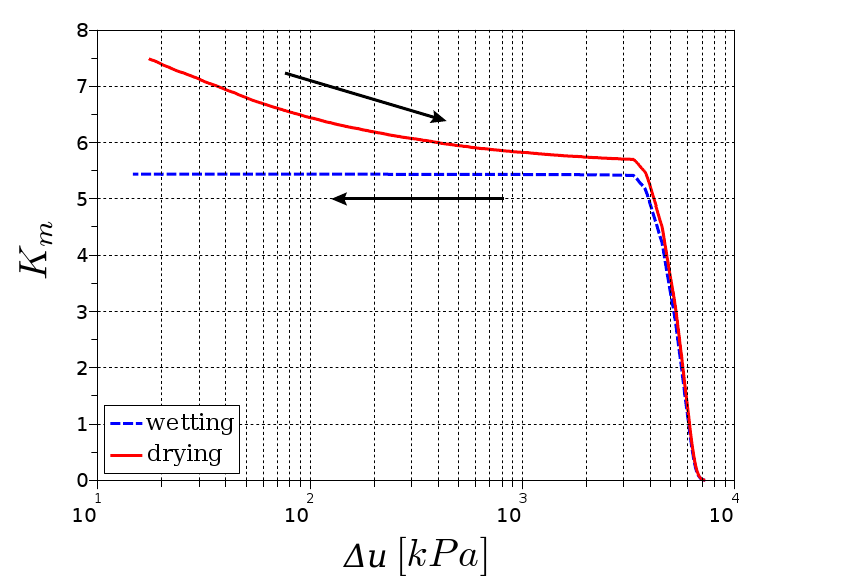}
\caption{Evolution of the average number of menisci per particle during both wetting and drying sequences.}
\label{HysteresisMenisciNumber}
\end{figure}

One can note that, through the possible local mechanisms of formation and breakage of menisci, an hydraulic hysteresis arises at the macroscopic level, depending on the wetting history of the material. This macroscopical hysteresis is directly linked to the liquid bridge density inside the sample. Indeed, due to the possibility of stretched liquid bridges between separated grains for a drying scenario, menisci are numerous inside the medium and the corresponding degrees of saturation are therefore larger for the simulated dried material than for the wetted one. \fig{HysteresisMenisciNumber} illustrates this point through the evolution of the average number of liquid bridge per particle $K_m$ during both the two scenarios.

It is remarkable that both curves present two distinct stages:
\begin{itemize}
\item [-] For $\Delta u > 3~MPa$, menisci distribution is driven by the particle gradation. Menisci progressively disappear with the increase in $\Delta u$ due to the persistence of capillary water between the smallest particles. For instance, this is the reason why clay needs higher capillary pressures than sand to be fully desaturated.
\item [-] For $\Delta u < 3~MPa$, menisci distribution is driven by the capillary pressure according to the wetting history of the material. During the drying scenario, the increase of $\Delta u$ induces smallest debonding distances (\fig{CapillaryPressureInfluence}) and therefore less stretched bridges inside the sample. During the wetting scenario, as menisci are only present at contacts, they keep constant in number whereas their respective volumes increase with the decrease of $\Delta u$.
\end{itemize}

At the particle scale, the simulated hysteresis can be illustrated through \fig{HysteresisScheme} where one can see that the darkest particle has more neighbours ($K_m = 5$) after a drying sequence, than after a wetting one ($K_m = 2$). This phenomenon, also denoted as the ink-bottle effect, is probably the most significant hysteretic mechanisms involved in unsaturated soil mechanics. The consequences of this property on the mechanical behaviour of granular assemblies is investigated in section 3.2.

\subsubsection{Stress-Strain response under triaxial loadings}
To determine the effect of interparticular capillary menisci on the mechanical properties of a granular material, we performed a series of triaxial compression loadings ($\sigma_2 = \sigma_3, \dot \epsilon_1 > 0$) on the numerical assembly under 5, 10 and 20~kPa confining pressures.

\begin{figure}
\centering
\includegraphics[width=\linewidth]{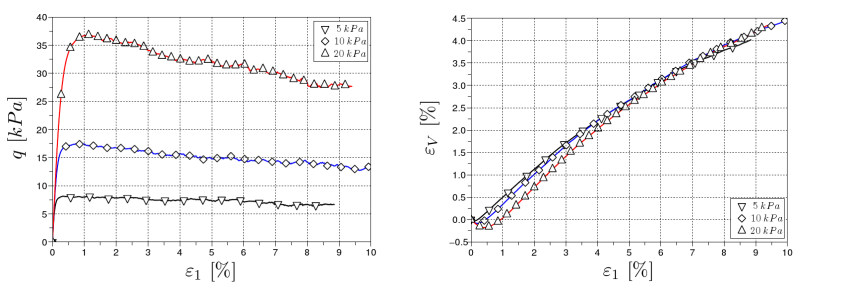}
\caption{Simulation of drained triaxial compressions on the dry specimen under 5, 10 and 20~kPa confining pressures.}
\label{QEpsVDry}
\end{figure}

As a reference, the response of the dry specimen is plotted in \fig{QEpsVDry}.
The assembly presents a behaviour which is qualitatively close to that of a dense granular material with strain softening associated with dilatancy.
\begin{figure}
\centering
\includegraphics[width=0.6\linewidth]{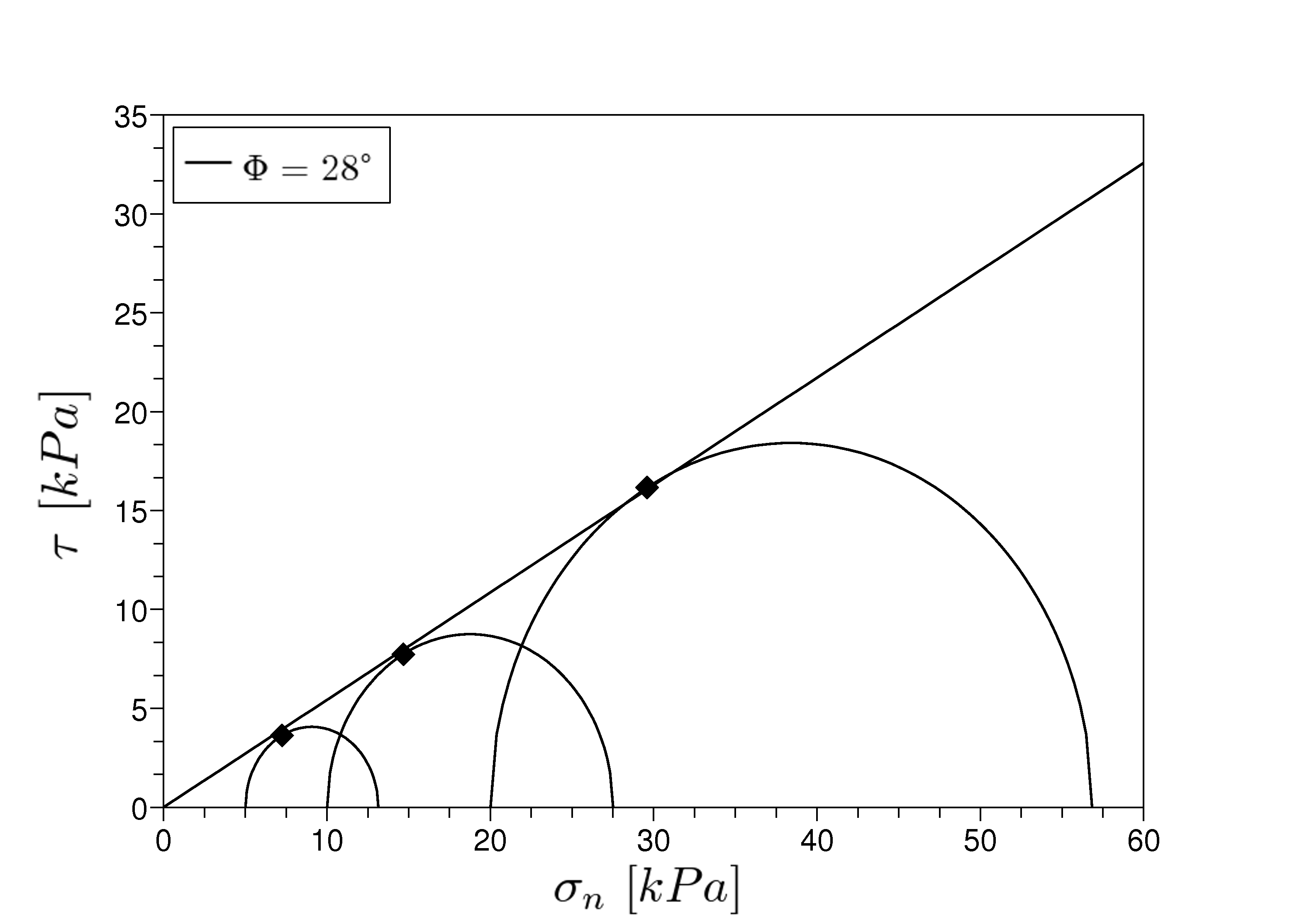}
\caption{Failure envelop for the dry sample.}
\label{MorhCoulombDry}
\end{figure}
As shown in \fig{MorhCoulombDry}, the discrete element model behaviour is well described through the Mohr-Coulomb criterion:
\begin{equation}
\tau = \sigma \; tan \, \Phi + c
\end{equation}
with a null apparent cohesion $c$ and an internal friction angle $\Phi$ of about
28$^{\circ}$.

In order to cover the entire pendular regime, numerical unsaturated triaxial tests were performed at different capillary pressures corresponding to degrees of saturation ranging from 0 to 10~\% along the wetting path. \tab{SrUc} presents the correspondence between $\Delta u$ and initial values of $Sr$ for the considered specimen.
\begin{table}[h]
\centering
\caption{Capillary pressures and corresponding initial degrees of saturation of the discrete element model}\label{SrUc}
\vskip1mm
\begin{tabular}{|l|c|c|c|c|c|c|}
\hline
Capillary pressure $\Delta u$~[kPa]	& $5000$ & $3000$ & $80$ & $50$ & $30$ & $20$ \\
\hline
Degree of saturation $Sr$~[\%]	& $0.001$ & $0.01$ & $1$ & $2.5$ & $5.5$ & $10$ \\
\hline
\end{tabular}
\end{table}

\begin{figure}
\centering
\includegraphics[width=\linewidth]{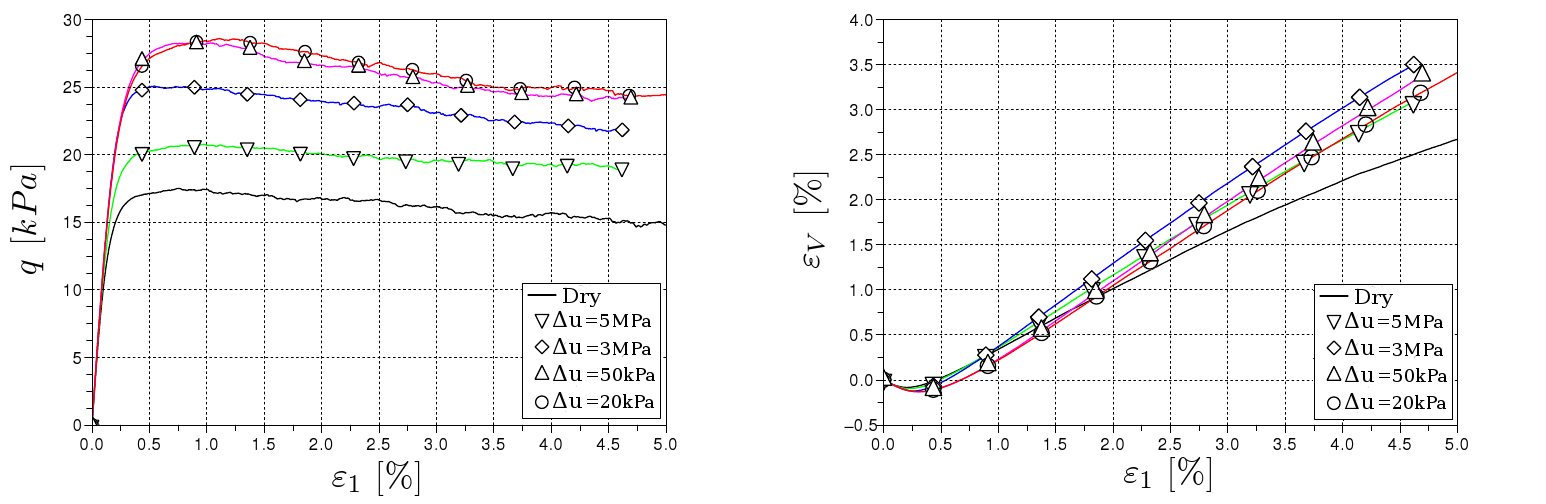}
\caption{Simulation of triaxial compressions at different capillary pressure levels under a 10~kPa confining pressure.}
\label{QEpsVSr}
\end{figure}

\begin{figure}
\centering
\includegraphics[width=\linewidth]{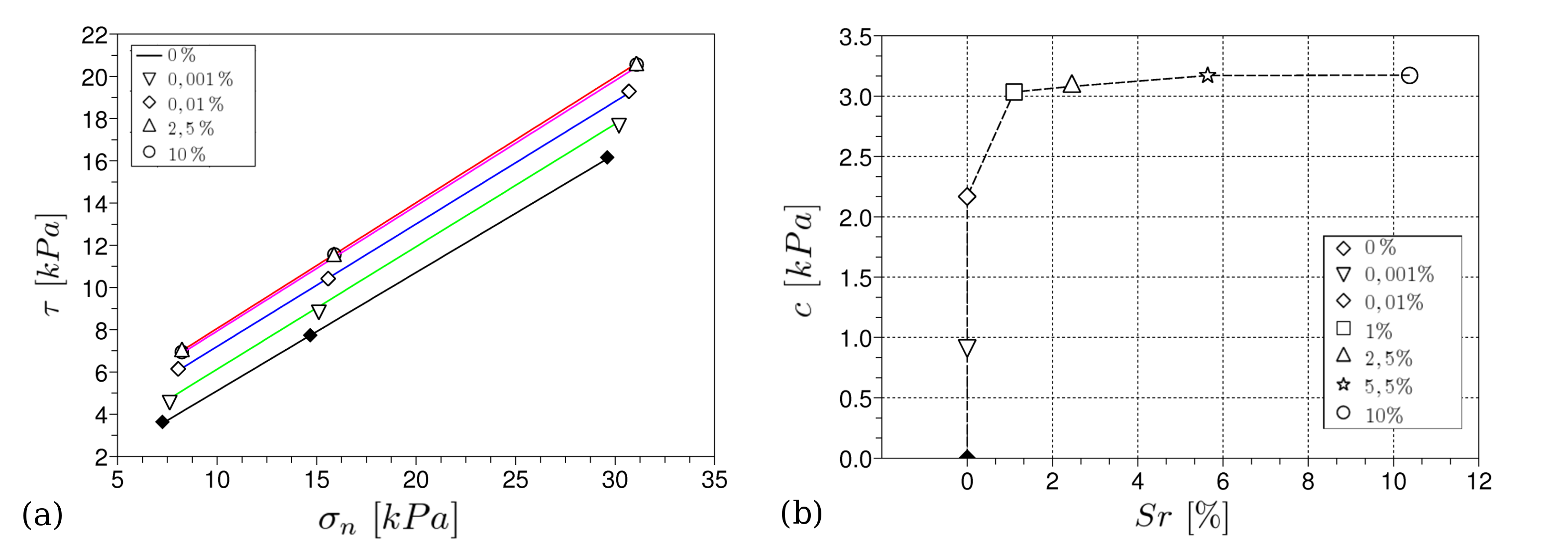}
\caption{(a) Morh-Coulomb failure envelops and (b) apparent cohesion $c$ as a function of the degree of saturation $Sr$.}
\label{MohrCoulombSrSr}
\end{figure}

\fig{QEpsVSr} displays the responses of the unsaturated samples compared to the dry one for triaxial compressions applied under a confining pressure of 10~kPa. For clarity reason, only four of the six tests are plotted. As expected, the shear strength is greater for unsaturated materials and depends on the degree of saturation. The more wetted the sample is, the higher the deviatoric strength is, with a trend for dilatancy more pronounced than in the dry case, suggesting a more interlocked structure. To sum up all simulated tests, the corresponding failure envelops as well as the evolution of the apparent cohesion $c$ with $Sr$ are plotted in \fig{MohrCoulombSrSr}.

As a first observation, one can see that the internal friction angle $\Phi$ is
independent on the degree of saturation, whatever the water content level. This
probably results from the assumption of null effect of capillarity on contact
friction, but it is remarkable that such a local property arises at the scale of
an assembly. \cite{Richefeu2006} came to the same conclusion from laboratory
experiments on glass beads assemblies with pure water, providing therefore a
good prediction to the idealized model. However, real granular materials such as
soil would certainly not highlight such an independence between local and global
friction properties due to the complex combined effects of surface roughness and
in situ liquid wetting features.

\begin{figure}
\centering
\includegraphics[width=0.5\linewidth]{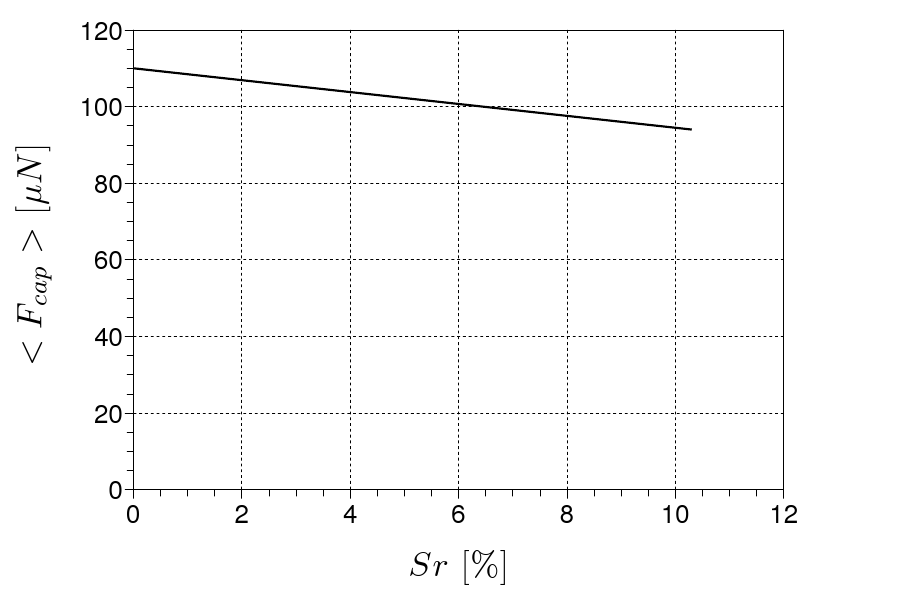}
\caption{Mean capillary force $< F_{cap} >$ as a function of the degree of saturation $Sr$.}
\label{FcapMoy}
\end{figure}

Second, it is clear that the cohesion varies significantly from dry to unsaturated states. $c$ increases non-linearly till a maximum value for the higher degrees of saturation that corresponds to the upper limit of the pendular state. This is fairly typical based on synthesis of reported experimental data on sand (\cite{Lu2007}) or glass beads (\cite{Richefeu2006}). Nevertheless, this is quite remarkable in regards to the interparticle behaviour, since capillary force intensity at contact is not such dependent on capillary pressure (\fig{CapillaryPressureInfluence}). For instance, computing the mean capillary force $< F_{cap} > = \frac{\sum^{N_m}_i F_{cap}^i}{N_m}$ in the specimen for all tested degrees of saturation (\fig{FcapMoy}) indicates that capillary forces tends to be greater, on average, for higher capillary pressure levels (and therefore for smaller saturation degrees).

This paradox is clearly due to capillary pressure influence on the debonding
distance of menisci as presented in \fig{CapillaryPressureInfluence}. Indeed,
strong internal rearrangements occur during the loading, leading to a
redistribution of the liquid inside the microstructure which is not obvious at
the macroscopic scale. How this redistribution occurs and how it influences the
shear strength of the material is strongly linked to the elaborated properties
of grains when they interact in an assembly. A micromechanical investigation is
therefore needed in order to identify effects induced by structural evolution of
the medium on the liquid bridge distribution.

\subsection{Microscopic analysis}
\begin{figure}
\centering
\includegraphics[width=\linewidth]{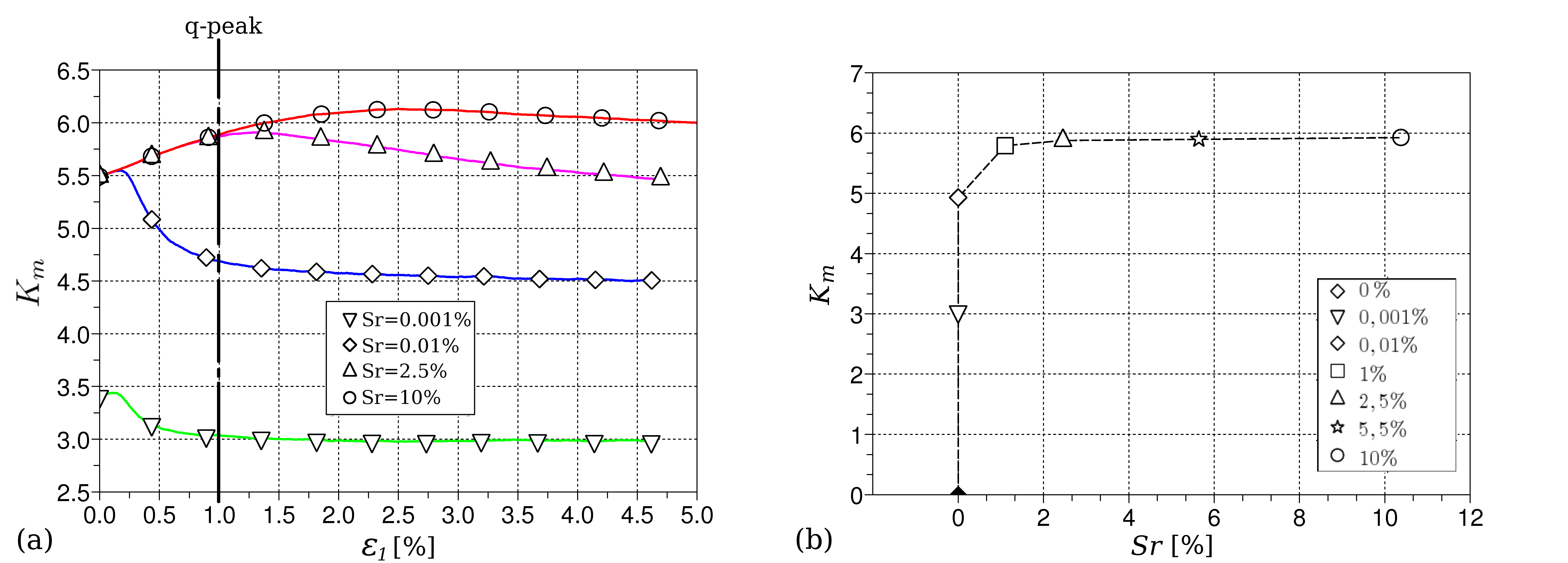}
\caption{(a) Evolution of the average number of menisci per particle $K_m$ during a triaxial compression for several initial degree of saturation $Sr$, (b) $K_m$ values at the q-peak as a function of $Sr$.}
\label{KmEps}
\end{figure}

\fig{KmEps}(a) shows the evolution of the average number of menisci per particle $K_m$ for different degrees of saturation during the unsaturated triaxial compressions presented before (\fig{QEpsVSr}).

Due to particle rearrangements during the loading, liquid bridge distribution inside the material changes with deformations, depending on the defined capillary pressure level. For instance, despite the fact that the initial liquid bridge densities are identical for tests corresponding to degrees of saturation of 0.01, 2.5, 10~\%, their evolutions during compression are totally different due to the local response of liquid bonds to loading. Typically, as a result of the local behaviour of menisci (see \fig{CapillaryPressureInfluence}), liquid bridges tend to persist more with the medium dilatancy for low values of $\Delta u$ (higher $Sr$), due to larger debonding distance $U_{rupture}$. As discussed in \cite{Scholtes2008} as well as in \cite{Scholtes2009}, the microstructure strongly influences liquid distribution inside the medium. The initial bridge density number appears therefore not appropriate to accurately evaluate the shear strength of the material as it was suggested by \cite{Richefeu2006}, in the sense that failure does not occur for an isotropic configuration of the material, but after some deformations leading to significant changes inside the fabric due to the induced anisotropy. Cohesion, as well as the internal friction angle, is dependent on space direction, and has to be evaluated at the yield state. Indeed, by plotting $K_m$ values corresponding to the $q-peak$ (\fig{KmEps}(b)), one can recover the shear strength hierarchy induced by capillarity depending on the degree of saturation.

It is important to note that, in the end, whatever the capillary pressure level or the degree of saturation, the pertinent parameter which determines the shear strength of a partially saturated granular material is the liquid bridge density inside the medium at the yield state. Obviously, the evolution of this density with deformations is totally driven by the capillary pressure level, and, thereafter, by the water content.

Another way to illustrate the primacy of the liquid bridge density on shear strength properties of a wet granular material is to compare its behaviour for both a drying and a wetting scenario. As suggested in section 3.1.1, a given sample can contain different numbers of capillary bonds for a defined $\Delta u$ value (\fig{HysteresisMenisciNumber}) due to the possible hydraulic hysteresis.

\begin{figure}
\centering
\includegraphics[width=\linewidth]{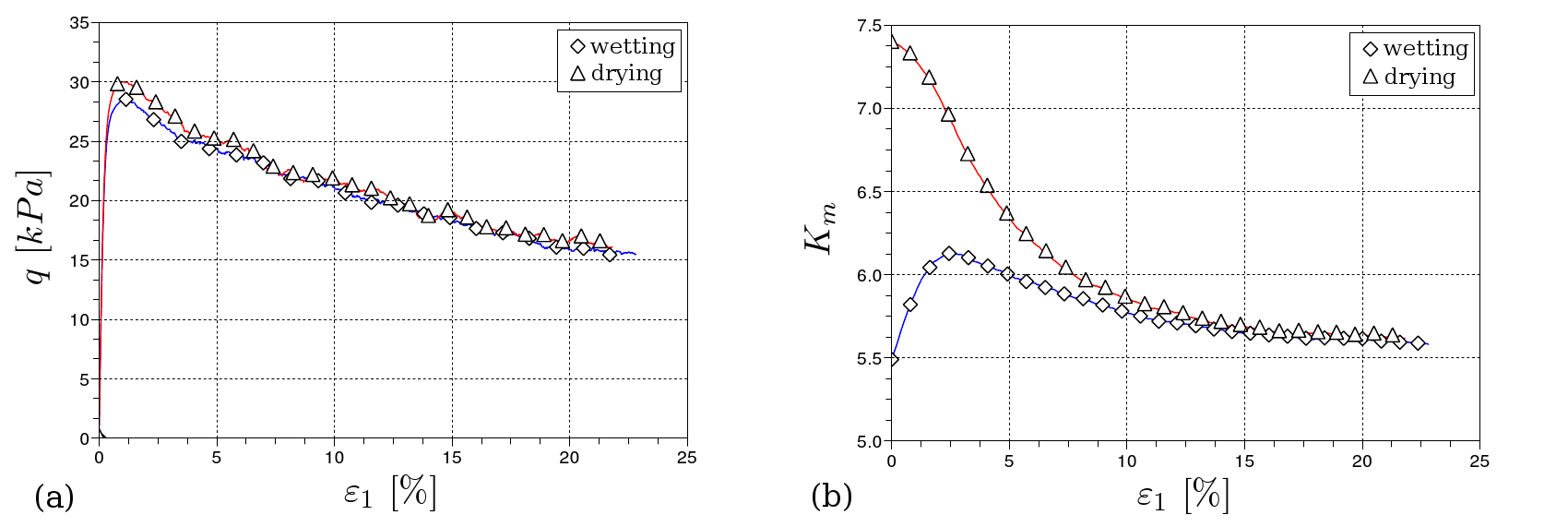}
\caption{(a) stress-strain relationships and (b) evolution of $K_m$ during a triaxial compression on a sample subjected to a drying and a wetting scenario respectively under the same capillary pressure $\Delta u$=20~kPa.}
\label{QKmVHysteresis}
\end{figure}

\fig{QKmVHysteresis}(a) presents the stress-strain relationships obtained for a numerical sample subjected to a triaxial loading under the same capillary pressure ($\Delta u$ = 20~kPa), but with two different initial configurations resulting respectively from a drying and a wetting scenario.
Due to numerous liquid bridges \fig{QKmVHysteresis}(b), the dried sample has a greater shear strength ($q_{peak}$=30~kpa) than the wetted one ($q_{peak}$=27~kpa), which results in respective cohesions of 5 and 6.5~kPa. It is remarkable that the ratio $\frac{c^{Drying}}{c^{Wetting}}=1.3$ is strongly correlated to the ratio between the corresponding values of $K_m$ at the $q-peak$: $\frac{K_m^{Drying}}{K_m^{Wetting}}=\frac{7.4}{5.5}=1.34$, confirming here the possibility to link $K_m^{q-peak}$ to the overall cohesion of the material.

Another remarkable feature of the coupled hydro-mechanical process is that internal rearrangements caused by the loading tend to conceal the initial difference between the two specimens, leading to a common residual state for large deformations with the same number of liquid bridges on average. This confirms the strong influence of the microstructure on the liquid phase distribution. For instance, if the presence of fluids inside a polyphasic granular material implies some consequences on its overall behaviour, the coupling is also active in the other way. Microstructural rearrangements of the solid skeleton lead to strong modifications of the fluid phases and both aspects induce complex mechanisms at the macroscopic scale.

\section{Conclusion}
A micromechanical computational model for the analysis of wet granular soils in the pendular regime has been proposed. Capillary mechanisms are described at the contact scale based upon the capillary theory in both terms of interparticle adhesive force and water retention. Capillary menisci are distributed inside the medium according to the thermodynamic equilibrium between liquid and gas phases. Moreover, an hydraulic hysteresis is accounted for based on the possible mechanisms of formation and breakage of liquid bridges during wetting and drying phases.

Triaxial compression test simulations were performed on a granular assembly
under several confining pressures for dry and partially saturated conditions in
order to analyze effects of local capillary menisci on macroscale cohesion and
friction properties. The results confirm that capillary-induced attractive
forces and hydraulic hysteresis play a significant role in the shear strength of
granular materials. By increasing normal forces at contact, capillary menisci
contribute to the apparent cohesion of the material and enhances its stiffness.
A remarkable aspect is that the shear strength of an assembly is basically
controlled by the liquid bridge density and not by the capillary pressure level.
A strong correlation is highlighted between the average number of liquid bridges
at the yield state and the induced cohesion. On the other hand, the liquid
bridge distribution inside the medium is controlled by the capillary pressure
which leads to counter-intuitive response at the scale of an assembly. In a
separate way to capillary forces, Coulomb cohesion increases with water content
and tends to a maximum value for the upper limit of the pendular regime
following the liquid bridge density. In additon, the results put in evidence the
effect of the skeleton induced anisotropy on liquid phase properties which is
generally occulted when considering hydro-mechanical modelling. Solid fabric
evolution clearly has a strong influence on the liquid distribution.

The proposed model is able to simulate the macroscopic response of wet granular materials and revealed a number of outstanding micromechanical mechanisms and response patterns consistent with experimental data. The multi-scale approach presented here appears to be a pertinent complementary tool for the study of unsaturated soil mechanics. A fundamental extension of this work would be to extend the interaction model to higher degrees of saturation with, for example, a validation through tomographic imaging.

\bibliographystyle{cmes}
\bibliography{ScholtesEtAl_CMES}


\lastpage

\end{document}